# Acoustic meta-stethoscope for cardiac auscultation

Zheng-Ji Chen, Jing-Jing Liu, Bin Liang*, Jing Yang & Jian-Chun Cheng

*Key Laboratory of Modern Acoustics of Ministry of Education, Collaborative Innovation Center of Advanced Microstructures, Department of Physics, Nanjing University, Nanjing 210093, China*

Corresponding author. E-mail address: liangbin@nju.edu.cn

**Abstract** Straight cylindrical stethoscopes serve as an important alternative to conventional stethoscopes whose application in the treatment of infectious diseases might be limited by the use of protective clothing. Yet their miniaturization is challenging due to the low-frequency of bioacoustics signal. Here, we design and experimentally implement a meta-stethoscope with subwavelength size, simple fabrication, easy assembly yet high sensitivity, which simply comprises multiple round perforated plate units and a cylindrical shell. We elucidate our proposed mechanism by analytically deriving the frequency response equation, which proves that the equivalent acoustic propagation path is substantially increased by the high-index metamaterial, enabling downscaling of the meta-stethoscope to subwavelength footprint. The acoustic performance of meta-stethoscope is experimentally characterized by monitoring the cardiac auscultation on clothed human body. The simulated and measured results agree well, with both showing the expected enhancement of sensitivity of our proposed meta-stethoscope (~10 dB) within the predicted working frequency range from 80 to 130 Hz despite its compactness and simplicity. Our designed portable, detachable yet effective meta-stethoscope opens a route to metamaterial-enabled stethoscope paradigm, with potential applications in diverse scenarios such as medical diagnosis and acoustic sensing.

Stethoscopes play an important role in clinical diagnosis for the advantages of economy, dynamic real-time monitoring and patient connection. However, the use of protective clothing substantially limits the application of traditional stethoscopes in many important scenarios such as infectious diseases like COVID-19 and H1N1[1-3]. To overcome this limitation, the design of a straight tube cylindrical stethoscope made of an empty potato chip tube is proposed which serves as a clinical diagnostic tool and an alternative to the traditional stethoscope [2], yet has an eigen-frequency much higher than the typical frequency of cardiac auscultation signals. Later Jiang et al. further optimize the configuration of such a simple cylindrical stethoscope by properly designing its diameter and length resulting in an improvement in sound intensity (3 dB higher than the original cylindrical stethoscope [4]. However, it still remains challenging to realize a portable structure given the long wavelength of heart sound signals. Given the trends towards miniaturization and integration, the mechanism for substantially downsizing the cylindrical stethoscope to portable scale while enhancing the auscultation sensitivity would be of fundamental interests and practical significance for many important applications ranging from medical diagnosis to acoustic sensing.

In this paper, we theoretically design and experimentally demonstrate a mechanism of a meta-stethoscope to break through this limitation. Although the recent emergence of acoustic metamaterials with special acoustical properties unattainable in the nature has enabled various wave-steering functionalities [5-8] such as sound absorption [9-11], diffuse reflection [12], among others, the application of metamaterials in the design of stethoscope still remains to be explored. Our designed meta-stethoscope simply comprises multiple layers of metamaterial unit cells implemented as a specially designed perforated round plate metamaterial which can be regarded as a dispersion-free structure below 300 Hz, the most significant maxima of the frequency components of the heart auscultation. The size of meta-stethoscope is deep wavelength (less than $\lambda/10$) that it can be carried everywhere, and meanwhile, through the appropriate selection of parameters and structure design, the meta-stethoscope can be disassembled or assembled. The theoretical results are in a good agreement with numerical results. That the results of auscultation experiment realize a high sound pressure level gain (10 dB) from 80 to 130 Hz compared to the original short tube confirms the effectiveness of the scheme. We use theoretical analyses, numerical simulations and experimental measurements to verify the effectiveness of our design bearing unique advantages of compactness, simplicity, portability, assemblability, and high sensitivity.

The schematic of our proposed design of meta-stethoscope is shown in Fig.1(a), which is composed of multiple layers of identical perforated plates periodically arranged in a cylindrical tube, with the overall length $L$, and the outer diameter of $D$. The diameter of circular hole on each plate is $R$, the distance between neighboring plate is $d$ and the plate thickness is $a$. All geometric parameters are much smaller than the wavelength of sound in air. The azimuth difference between two adjacent units is $\pi$. Instead of adopting the previous metamaterial designs such as coiling-up space structure [15-22,24], tapered labyrinthine structure [19,23] and helical structure [25], here we propose such a compact and simplest possible design of metamaterial unit cells which can be easily fabricated at a low cost and conveniently disassembled/assembled. This is ensured by the fact that the cylindrical shell is divided into two parts along the axis, and etched with thirteen grooves (with the width equal to the thickness of the plate) on the inner wall. The shape of each groove is exactly the same with depth and width both being 1 mm. Then we insert the perforated plate units into the grooves and assemble together with the cylindrical shell to form a meta-stethoscope. As shown in the schematic diagram in Figs.1(a) and (b), the incident waves from the patient's chest are forced to propagate through the holes in circular plates and then enter the outlet at the other end. As a consequence of such a substantial increase of the propagation distance and the associated local resonance, a high refractive index and the negligible transmission loss of perforated plates are enabled at very low frequency regime which are responsible for the unique advantages of the meta-stethoscope in terms of sensitivity and compactnesss [27], as will be demonstrated both theoretically and experimentally.

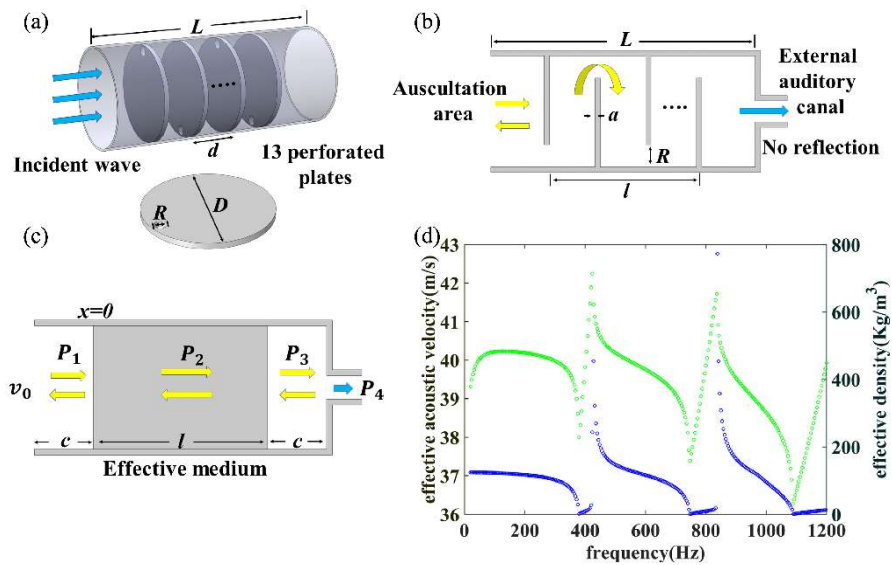

**FIG.1 | Schematic of the proposed Meta-stethoscope and the calculated effective material properties. (a)** A cylindrical meta-stethoscope unit consists of a cylindrical tube which is divided into two semi-cylindrical shells and thirteen perforated plate units, the geometry of which is determined by a length $L=138$ mm, $l=98$ mm, an outer diameter $D=42$ mm, a diameter of circular hole $R=4.2$ mm, a distance between centers of the two circles $d=6$ mm and a thickness $a=2$ mm. **(b)** The schematic diagram of the meta-stethoscope. **(c)** The schematic diagram of equivalent medium. **(d)** The effective acoustic velocity $c_{\text{eff}}$ (the green line) and dynamic mass densities $\rho_{\text{eff}}$ (the blue line) change with the frequency of the meta-stethoscope, calculated by Equation.

First we begin from the theoretical study of the sound propagation in the proposed meta-stethoscope. Due to deep-subwavelength nature of the designed metamaterial unit cells, the meta-stethoscope can be treated as a homogeneous medium characterized by effective acoustical parameters. From the effective medium point of view [18,25], the perforated metamaterial may behave as a homogeneous column. We analytically derive the effective parameters as follows:

$$c_{eff} = 2\pi f l / \sin^{-1}(\sqrt{-t_{12}t_{21}}), \rho_{eff} = \sqrt{t_{12}/t_{21}}/c_{eff} \qquad (1)$$

where $t_{12}$, $t_{21}$ are components of the transfer matrix for the metamaterial layer and f is the frequency of acoustic waves in air.

The theoretical results of the effective sound speed and dynamic mass density are depicted in Fig. 1(d). It is observed that despite the obvious dispersion caused by the resonance nature, the metamaterial exhibits a substantial constant reduction in the effective velocity (as low as 40 m/s) within a relatively broad low-frequency range, which will be proven crucial for the downscaling of the structure and improvement of its sensitivity.

Based on the equivalent parameters above, the acoustic properties of meta-stethoscope can be characterized by using acoustic waveguide and equivalent medium theory. As is shown in Fig.1(c), there is an air gap of 2 cm between two ends of the probe and the adjacent perforated plate unit for practical application. We divide the sound field into four parts, and there is only plane wave mode inside the meta-stethoscope for the working frequency is lower than the cutoff frequency. The sound pressure and vibration velocity in different sound regions can be expressed as:

$$p_n = A_n e^{-ik_n x} + B_n e^{ik_n x}, v_n = (A_n e^{-ik_n x} - B_n e^{ik_n x})/(\rho_n c_n) \qquad (2)$$

Where $n$=1, 2, 3, 4, $k_n$=$2\pi f/c_n$, $c_{1,3,4}$ is the velocity of sound in air, $c_2$ is the velocity of sound in effective medium, $\rho_{1,3,4}$ is the density of air, $\rho_2$ is the dynamic mass density of effective medium. By

invoking the the boundary conditions and the continuity between sound pressure and volume velocity:

$$x = -c, v(-c) = v_0, (A_1 - B_1)/\rho_1 c_1 = v_0 \tag{3}$$

$$x = 0, A_1 + B_1 = A_2 + B_2, (A_1 - B_1)\rho_2 c_2 = (A_2 - B_2)\rho_1 c_1 \tag{4}$$

$$x = l, A_2 e^{-ik_2 l} + B_2 e^{ik_2 l} = A_3 e^{-ik_1 l} + B_3 e^{ik_1 l}$$
$$(A_2 e^{-ik_2 l} - B_2 e^{ik_2 l})/\rho_2 c_2 = (A_3 e^{-ik_1 l} - B_3 e^{ik_1 l})/\rho_1 c_1 \tag{5}$$

$$x = l + c, A_3 e^{-ik_1(l+c)} + B_3 e^{ik_1(l+c)} = A_4 e^{-ik_1(l+c)}$$
$$S_0 (A_3 e^{-ik_1(l+c)} - B_3 e^{ik_1(l+c)}) = S_1 A_4 e^{-ik_1(l+c)} \tag{6}$$

From equations above, with the help of transfer matrix, we obtain:

$$SPL = 20\log_{10}|A_4|/\sqrt{2}/P_{ref}, A_4 = 2\rho_1 c_1 v_0 e^{ik_1(L+l)}/E \tag{7}$$

$$E = i[2\cos(k_2 L)\sin(2k_1 l) + (z_{12} + z_{21})\sin(k_2 L)\cos(2k_1 l) + (z_{12} - z_{21})$$
$$\times \sin(k_2 L)] + 2S_1 \cos(k_2 L)\cos(2k_1 l)/S_0 - S_1(z_{12} + z_{21})\sin(k_2 l)\sin(2k_1 l)/S_0 \tag{8}$$

The analytical expression of frequency response given in Eq. 7 clearly suggests a remarkable enhancement of sensitivity as the frequency approaches heart sound band, which is much lower than the resonance frequency of a cylindrical stethoscope of equal length and can cover the typical heart signal band (~200 Hz) with a portable length (~13 cm). As a result, our proposed meta-stethoscope is expected to enable the desired heart auscultation with high sensitivity by using a portable-sized structure, as will be verified in what follows. It is worth noting that when the equivalent medium is air instead of the metamaterial, the stethoscope is indeed an original straight tube with length $L$.

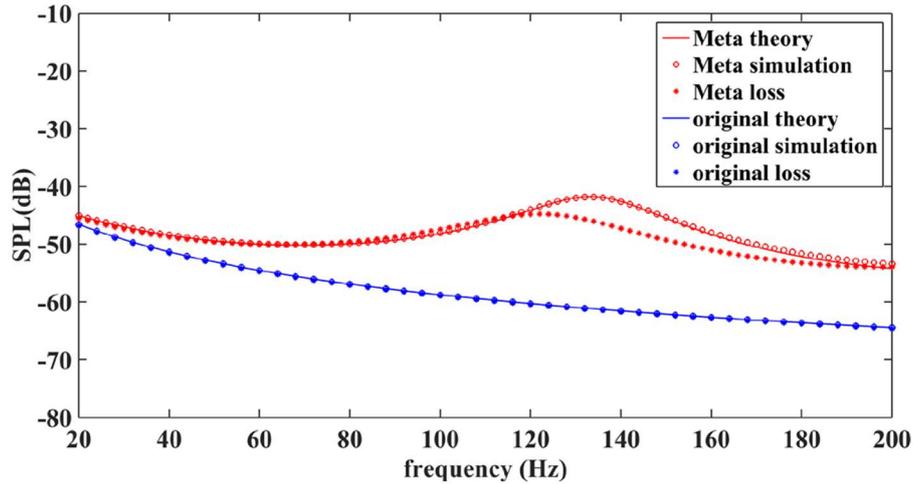

**FIG.2 | Theoretical and simulated transmission spectra for the meta-stethoscope and original tube.** The red(blue) line, red(blue) circle and red(blue) star separately represent theory, no-loss simulation and thermo-viscous simulation of the meta-stethoscope (original straight tube)

The comparison between the results of analytical prediction and numerical simulations is shown in Fig. 2, where a case of cylindrical stethoscope is also included. In this work, the simulation was carried out by using COMSOL MULTIPHYSICS$^{TM}$. A nearly-perfect agreement between the theoretical and numerical results is observed which both show the expected acoustical performance featured a high sensitivity.

All data is in a great agreement except the loss spectrum of meta-stethoscope, which confirms the rationality of the equivalent medium theory and the correctness of the results. There is almost no loss for original tube because of the large diameter and no formant in the frequency band. In contrast, the formant of the meta-stethoscope drops from 132 Hz in theory to about 120 Hz and the peak value also decreases due to the thermo-viscous loss, but the sound pressure level is still significantly higher than that of the original tube. In particular, there is a gain of more than 10 dB at the peak against the cylindrical stethoscope. The results above show that meta-stethoscope can indeed lower resonant frequency and increase the auscultation sensitivity.

Next we verify the effectiveness of our mechanism via experimental measurements of the performance of the designed meta-stethoscope. The experiments were conducted in the anechoic chamber for the meta-stethoscope and an original stethoscope. All samples were fabricated via a 3-D printing technique (Stratasys Dimension Elite, 0.177 mm in precision), made of acrylonitrile butadience styrene (ABS) plastic, with the density and sound velocity being 1230 kg/m$^3$, and 2230 m/s respectively, which can be regarded as acoustically rigid due to the huge impedance contrast between solid and air. The transmission spectra were measured by using a 1/4-in.-diameter Brüel & Kjær type-4961 microphone. The frequency band of test is from 20 Hz to 200 Hz. The tests were performed at one-minute interval and each lasted for 30 seconds, with a layer of underwear being between the stethoscope and the volunteer's body. At least eight groups of measurements were separately implemented for the meta-stethoscope, the original tube and ambient noise, and the average was taken to obtain the average sound pressure level.

The measured amplitude spectra of the sound pressure signals are presented in Fig. 3(a). The comparison of ambient noise and measured sound pressure levels shows that the meta-stethoscope can be used for heart auscultation because the sound pressure levels of both stethoscopes are at least 10 dB higher than that of ambient noise. The experimental amplitude spectrum of the meta-stethoscope decreases with the frequency at lower and higher frequency band, while an obvious formant appears in the middle of the frequency band and the sound pressure levels are significantly higher than that of the original tube, which is consistent with simulation results. In stark contrast, there is an obvious peak in the frequency response of our meta-stethoscope, which amplifies the sound intensity of the received acoustic wave at low-frequency range unattainable with the original design. As a result of such amplification effect, the meta-stethoscope improves the average sensitivity by approximately 10 dB within the typical cardiac sound frequency range while reducing the overall device length to 1/13 $\lambda$, which is important for its practical application. The measured peak is slightly lower than the numerical results, which is primarily due to the imperfect fabrication and alignment of sample that leads to additional viscosity and leaking of acoustic energy. This further manifests the unique advantage of our designed meta-stethoscope in terms of simplicity and portability, as evidence by the satisfying acoustical performance even in the presence of the above experimental errors in such a manually-assembled device.

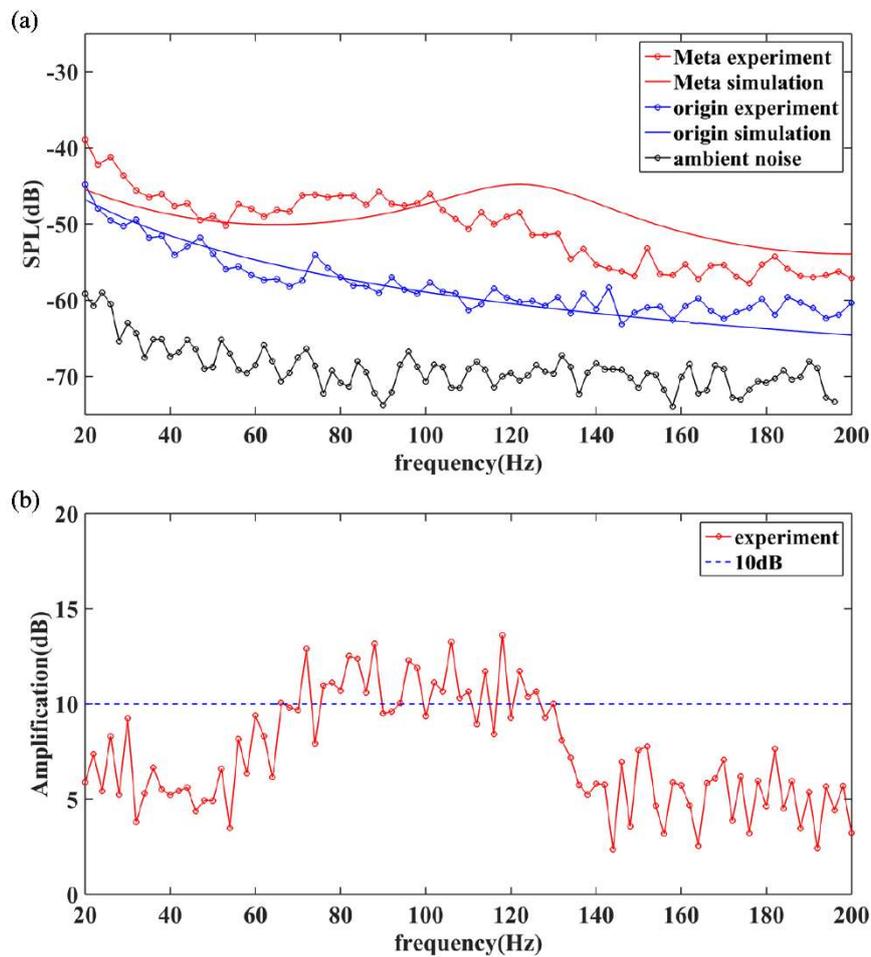

**FIG.3 | The frequency spectra of sound pressure levels measured in an anechoic chamber. (a)** Numerically calculated and experimentally measured transmission spectra for meta-stethoscope , original tube and ambient noise. **(b)** The amplification of the meta-stethoscope against the original tube.

For a quantitative estimation of the gain of sensitivity, we introduce a parameter for the meta-stethoscope, defined as amplification, and plot the results in Fig. 3(b). There is a gain of more than 10 dB between 80 and 130 Hz, more than 4 dB when less than 80 Hz and 3 dB when more than 140 Hz, which verifies that the meta-stethoscope possesses better auscultation effect in the typical heart sound frequency band.

In summary, we have proposed and experimentally demonstrated a meta-stethoscope based on a perforated round plate metamaterial, which realizes better auscultatory performance than the original stethoscope in heart sound frequency band, especially an amplification of 10 dB between 80 and 130 Hz with the deep sub-wavelength diameter and length. Although the perforated metamaterial shows dispersive acoustic impedance over a broad frequency band, it remains non-dispersive below 300Hz, which allows us to use the equivalent medium method for theoretically research. Such remarkable characters derive from the high-refractive-indexed metamaterial and resonance effects inside the stethoscope. The meta-stethoscope shares the advantages of compactness and portability. It can be used as a diagnostic tool for medical workers at risk of infection who need to wear protective clothing. Our work is the first attempt to apply acoustic metamaterials to stethoscope and another demonstration of the large potential of acoustic metamaterials, which will help promote the application of metamaterials in the field of medical diagnosis.

**ACKNOWLEDGEMENTS**

*This work was supported by the National Key R&D Program of China (Grant Nos. 2022YFA1404402 and 2017YFA0303700), the National Natural Science Foundation of China*

*(Grant Nos. 11634006 and 12174190), High-Performance Computing Center of Collaborative Innovation Center of Advanced Microstructures and a project funded by the Priority Academic Program Development of Jiangsu Higher Education Institutions.*

**AUTHOR DECLARATIONS**

**Conflict of Interest**

The authors have no conflicts to disclose.